# PTP: Path-specified Transport Protocol for Concurrent Multipath Transmission in Named Data Networks

Yuhang Ye[†], Brian Lee[†], Ronan Flynn[†], Niall Murray[†], Guiming Fang[‡], Jianwen Cao[‡], Yuansong Qiao[†]

[†]Software Research Institute, Athlone Institute of Technology, Athlone, Co. Westmeath, Ireland
[‡]Institute of Software, Chinese Academy of Sciences, Beijing, China
{yye, ysqiao, nmurray}@research.ait.ie, {blee, rflynn}@ait.ie, {fgm, jianwen}@iscas.ac.cn

## ABSTRACT

Named Data Networking (NDN) is a promising Future Internet architecture to support content distribution. Its inherent addressless routing paradigm brings valuable characteristics to improve the transmission robustness and efficiency, e.g. users are enabled to download content from multiple providers concurrently. However, multipath transmission NDN is different from that in Multipath TCP, i.e. the "paths" in NDN are transparent to and uncontrollable by users. To this end, the user controls the traffic on all transmission paths as an entirety, which leads to a noticeable problem of low bandwidth utilization. In particular, the congestion of a certain path will trigger traffic reduction on other transmission paths that are underutilized. Some solutions have been proposed by letting routers balance loads of different paths to avoid congesting a certain path prematurely. However, the complexity of obtaining an optimal load balancing solution (i.e. solving a Multi-Commodity Flow problem) becomes higher with increasing network size, which limits universal NDN deployments.

This paper introduces a compromise solution – Path-specified Transport Protocol (PTP). PTP supports both the label switching and the addressless routing schemes. Specifically, the label switching scheme allows users to precisely control the traffic on each transmission path, and the addressless routing scheme maintains the valuable feature of retrieving content from any provider to guarantee robustness. As the traffic on a transmission path can be explicitly controlled by consumers, load balancing is no longer needed in routers, which reduces the computational burden on routers and consequently increases the system scalability. The experimental results show that PTP significantly increases users' downloading rates and improves network throughput.

## 1. Introduction

During the last decade, content-oriented applications have dominated Internet usage. Aiming to enhance content distribution performance, Information-Centric Networking (ICN) is proposed. NDN [1] is an effective ICN architecture based on the principle that the request/content packets are delivered according to a content-related universal resource identifier (name) without specifying the host addresses of content owners. The content retrieval in NDN is initiated by issuing requests (Interest packets) from users (consumers). An Interest packet is forwarded to a trusted content provider (a producer or cache) by a name-based routing scheme. Afterwards, the resulting Data packet follows the reverse path taken by the Interest packet and is delivered to the consumer. As NDN removes the address information from packets, content can be downloaded from any content providers.

The NDN's addressless name-based routing scheme results in unique traffic characteristics. The transmitting traffic that aims to deliver a content object is defined as a flow in this paper. A content object can be a whole file or a large chunk (e.g. video chunk) that can be further sliced into Data packets. Traffic is naturally receiver-driven [1], where consumers send Interest packets to pull Data packets from producers. A content object can be disseminated to multiple locations within the network, which enables consumers to concurrently download different parts of the same object from different providers, which is also known as multisource transmission. Additionally, the transmission path between a single consumer and a single content provider may not be unique, resulting in multipath transmission. As a content object can be provided by any authorized producer via any eligible path, the failure of a single producer or a single path will not interrupt the downloading, which improves the robustness of transmission. In addition, by utilizing the bandwidth of multiple paths/sources concurrently, the consumer's downloading rate can be improved. Nevertheless, the name-based routing scheme also brings challenges in traffic control. Under the TCP architecture, the traffic is controlled for each end-to-end flow. However, as NDN lacks the concept of connection, it is not feasible to control the flow on a single transmission path (from one producer to one consumer). Instead, the consumer treats the flow on all transmission paths (from multiple producers to one consumer) as an entirety and controls it using a single-path congestion control law. Under this design, if one of the transmission paths is congested before the others, the consumer will face a dilemma – "whether or not to reduce the requesting rate?" For minimizing congestion, the answer is *YES*, but for maximizing throughput, the answer is *NO*.

An intuitive solution to this dilemma, unbalanced utilizations on different paths, is to design a forwarding strategy that can balance the load on each transmission path by routers. As the optimal traffic allocation for multiple flows in a network can be viewed as solving a Multi-Commodity Flow problem [2], the computation and communication overheads for a distributed algorithm becomes high in large-scale networks. Although some approximate models have been proposed to utilize local measurements (round-trip time and pending Interests) for load balancing [2], [3], the experimental results [4] show that they may cause load imbalance on different paths in certain scenarios and degenerate bandwidth utilization when cooperating with receiver-driven congestion control.

Instead of preventing the dilemma, this paper starts from the different perspective of removing the dilemma. The proposed solution, Path-specified Transport Protocol (PTP), enables consumers to control the traffic on each transmission path explicitly. The two sub-protocols are included in PTP, namely, Multipath Path-specified Forwarding Protocol (MPFP) and Multipath Path-specified Congestion Control Protocol (MPCCP).

In MPFP, the consumers specify the transmission path of each Interest packet via a tag (path label), and the routers follow the tag's instruction to forward the Interest packet to the specified interface. It may be argued that pre-defined paths limit the forwarding dynamics provided by the original name-based routing, which is against NDN's goal and negates some advantages (e.g. robustness). To retain the advantages and also obtain the ability to control the traffic on each path, MPFP is a hybrid solution. It applies the path-specified forwarding scheme preferentially if the transmission path is connected and stable; otherwise, the name-based routing scheme is applied. The advantage brought by MPFP is its reduced computational requirement at routers. As the hop-by-hop forwarding interface (transmission path) of an Interest packet is determined by consumers, routers are only responsible for judging the validity of the specified forwarding interface but do not need to balance the load. In addition, MPFP introduces an efficient Forwarding Acceleration Base (FAB) for elephant flows, which allows bypassing the costly Longest Prefix Match (LPM) procedure, and consequently reduces the routers' FIB lookup overheads.

As MPFP separates the traffic control for each path, the traffic optimization objective becomes very similar to that in MPTCP [5]. MPCCP aims to control the amount of inflight packets on multiple paths to balance multi-flow fairness [6] and maximize bandwidth utilization with a bounded congestion level. Thus, an off-the-shelf MPTCP congestion control law becomes suitable. In order to support the NDN's special traffic characteristics (e.g. in-network caching and traffic aggregation) that affect measuring congestion, MPCCP integrates the MPTCP congestion control law with a proposed congestion detection method that distinguishes the content providers (e.g. cache or server) on the same path.

Comparing PTP with state-of-the-art transport protocols (e.g. receiver-driven, hop-by-hop and hybrid methods shows a significant improvement in terms of increasing throughput and enhancing downloading rates. The key contributions of this paper include:

- Separated congestion control for each transmission path to improve overall bandwidth utilization.
- Offloaded computational load from routers to consumers by determining the transmission path at the consumers.
- Decreased cost of forwarding table look-up through replacing Longest Prefix Match by an exact match.
- Combination of a proposed novel congestion detection scheme with an MPTCP congestion control law to support in-network caching and traffic aggregation in NDN.

The rest of the paper is structured as follows. Section 2 discusses the problems in multisource/multipath forwarding. Section 3 introduces the architecture overview of PTP. Section 4 and Section 5 illustrate the detailed implementation of PTP. Section 6 demonstrates the effectiveness of PTP through comparisons with state-of-the-art transport protocols. Section 7 gives an overview of the related works. Section 8 concludes the paper.

## 2. Background and Motivation

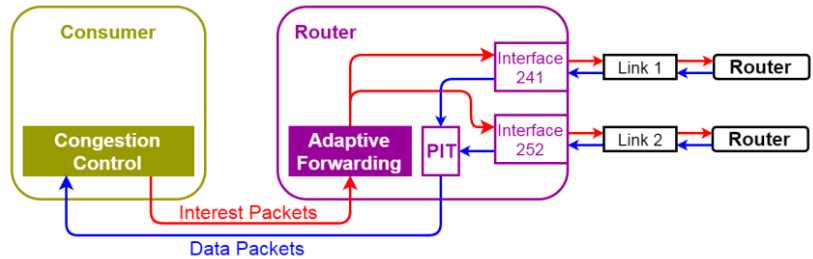

Figure 1 Conventional Transport Framework

As the Interest forwarding/routing in NDN is based on prefix, and each prefix may correspond to multiple content producers, NDN can natively support multisource/multipath transmission. Under this design, the overall downloading rate of a consumer is the summation of the downloading rates of all in-use paths, and the traffic on each link is the summation of the requested traffic by all consumers. To utilize the multipath bandwidths provided by the network without congestion, the conventional transmission framework in NDN is composed of two functional modules. They are congestion control and adaptive forwarding (Figure 1), where receiver-driven congestion control is deployed at consumers and hop-by-hop adaptive forwarding is deployed at routers.

In general, the two functions jointly determine the traffic on each link. Due to the fact that NDN forwards (anycasts) Interest packet according to its prefix, the consumer cannot distinguish or control the traffic on each transmission path (to content) independently. Thus, the consumer has to assume the traffic that comes from the network as an entirety and must control it as an entirety. To distribute the traffic (required by the consumer) to the suitable transmission paths, adaptive forwarding enables routers to split traffic to different interfaces/links. The majority of the NDN transport approaches fall into this framework, and the main advantage of this framework is that strict end-to-end communication is entirely avoided for packet forwarding, which consequently increases flexibility and robustness of packet transmission. In addition to the conventional framework, NDN also supports in-network congestion control, which is known as hop-by-hop Interest shaping. The pros and cons of two frameworks will be further discussed in the following section.

*A. Multipath Transmission Challenge in the Conventional Framework*

From the perspective of maximizing bandwidth utilization, the ideal traffic allocation to each link can be considered as a multi-commodity flow problem [2] that is to precisely fill all bottlenecks (minimum cut links). Under the conventional framework, the traffic allocation to a link is decided by both congestion control (origin of the traffic) and adaptive forwarding (split of the traffic). The congestion happens when the traffic on a bottleneck exceeds its capacity, which may be caused by consumers requesting excessive content (Data packets) from the producers and consequently overloading the bottlenecks. To prevent intensifying congestion, a straightforward solution is to notify consumers (e.g. via ECN [7], congestion NACK [8] and latency-based congestion detection [2]) and let them reduce the requesting rates. In addition to minimizing congestion, a major benefit is that it guarantees the consumers' low loss rate, e.g. consumers reduce the requesting rate to decrease the packet loss rate and the queuing latency.

In addition to the case when consumers exhaust the network resources, the routers' improper forwarding allocations may also cause congestion before exhausting bandwidth. The rationale is that the solution of the Multi-commodity flow problem assumes that all the bottlenecks are precisely filled at the **same** time, which is not always feasible in a real network. As existing solutions do not consider cooperation between routers and consumers (congestion control and adaptive forwarding are functionally independent), if the routers allocate excessive traffic to a certain bottleneck and overload it earlier than the others, consumers have to reduce the requesting rates to prevent exacerbating congestion and reduce loss rate. This results in the underutilization of the other congestion-free paths that are never fully filled.

*B. Multipath Transmission Challenge in the Hop-by-hop Framework*

The hop-by-hop framework (e.g. Interest shaping scheme [9]) can prevent congestion at routers by controlling the Interest forwarding rate without the support of consumers. Although this framework prevents the congestion caused by Data packets, it still requires the consumers' congestion control to satisfy the requirements for different applications. For example, suppose consumers continuously request a huge amount of Data packets that can fill all bottlenecks and at the same time each router proactively discards excessive Interest packets to minimize queuing latency, the consumers will achieve high bandwidth utilization and low packet delivery latency. However, the packet loss rate will be high. In the case that consumers and routers jointly control the traffic on each link, the problem of the conventional framework still exist. This is because routers still need to estimate optimal forwarding allocations to different interfaces to balance the load on different paths, which is identical to the conventional framework.

*C. A State-of-the-art Example: OMCC-RF*

This section uses an example to show why an inaccurate forwarding allocation may degenerate the bandwidth utilization. Optimal Multipath Congestion Control and Request Forwarding (OMCC-RF) [2] is a state-of-the-art transport solution that belongs to the conventional framework and is based on joint optimization. The adaptive forwarding strategy in OMCC-RF, Request Forwarding Algorithm (RFA), considers minimizing the unbalanced traffic at routers to balance the traffic to different bottlenecks, where the unbalanced traffic is approximated by the number of pending Interest (PI) packets, which are the packets that have been sent but not received at each forwarding interface. The rationale is that PI reflects the cost of the following downstream paths, the increasing number of PI indicates a congestion event at the downstream paths. Specifically, the number of PI packets is an approximate metric that reflects the relative congestion level if the characteristics (e.g. the inherent network latencies and bandwidths) of different transmission paths are very similar. In this case, the maximal number of in-transmission packets in a congestion-free path is its Bandwidth Delay Product (BDP). The number of PI packets of a path is the summation of its BDP and the number of the congested packets. As BDPs are similar for different paths, a higher PI value of a path indicates a heavier congestion.

However, the relationship between the number of PI packets and the relative congestion level of a path is more complex. As the processing delays, the end-to-end propagation delays, the multi-hop transmission delays and the bottleneck bandwidths may be very different to the heterogeneous paths, it is inaccurate to decide that the path with a smaller number of PI packets is less congested than the one with a larger number of PI packets. The critical issues of OMCC-RF are presented using the following example.

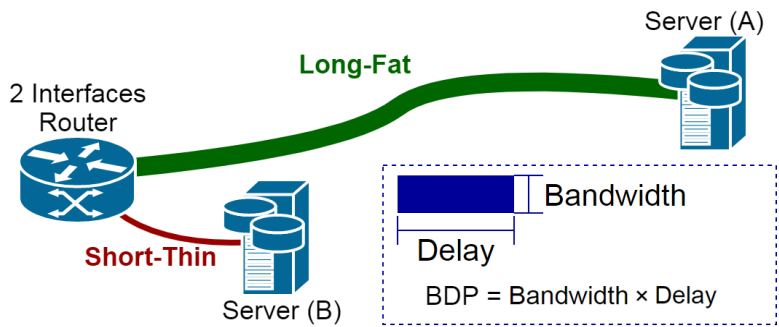

Figure 2  A Router Connected to Two Producers

As shown in Figure 2, a router is equipped with two interfaces, one is connecting to a long-fat path and the other one is connecting to a short-thin path. The long-fat path requires a large amount of inflight packets (i.e. equal to BDP) to achieve the optimal utilization and consequently indicates a large number of PI packets, while the short-thin one only requires a small amount. By deploying the RFA strategy at the router, a noticeable time point is when both the long-fat path and the short-thin path are both filled with a number of packets that equals to the short-thin path's BDP $w_{st}$. In this case, the short-thin path is fully filled. If the consumer keeps increasing the requesting rate, the queuing packets will be built up at the bottleneck of the short-thin path, which results in congestion at this path. Let the number of the queuing packets at the short-thin path be $q_{st}$, the number of PI packets in the two paths will be balanced at $w_{st} + q_{st}$. If $q_{st}$ satisfies the requirement in eq. (1), the long-fat path can be fully utilized.

$$w_{lf} \leq w_{st} + q_{st} \Rightarrow q_{st} \geq w_{lf} - w_{st} \tag{1}$$

where $w_{lf}$ denotes the BDP of the long-fat path. The rationale of eq. (1) is that when the number of the inflight packets of either the long-fat path or the short-thin one is no smaller than the corresponding BDP, the bandwidths of the two paths are fully utilized. However, the OMCC-RF's congestion control scheme – Remote Adaptive Active Queue Management (RAAQM) detects congestion by monitoring the variation of the queuing delay of **each** path, which usually targets minimizing the number of queuing packets, i.e. $q_{st}$ is very small. In this case, the number of queuing packets at the short-thin path $q_{st}$ is less like to satisfy the requirement of eq. (1), thus the bandwidth of the long-fat path is underutilized. It may be argued that if RFA is cooperating with another congestion control scheme that can satisfy the requirement of eq. (1), i.e. it enables the short-thin path to backlog sufficient Data packets, the utilization can be improved. However, the large queuing delay may affect the Quality of Experience (QoE) [10] of some real-time applications. For instance, the user will suffer from large end-to-end latencies during the video conferencing if the queuing delay is large. Moreover, the overly large queue may cause bufferbloat and exacerbate the instability of the network [11]. It is also worth noting that the latency-based metric (e.g. RTT) suffers from this utilization issue as well. The rationale is that the corresponding strategy prefers to overload the low latency path earlier than the high latency paths.

*D. Summary and Motivation*

According to our observations, the existing solution in hop-by-hop adaptive forwarding is not able to maximize the bandwidth utilization of the network. The rationale is that the conventional framework prevents consumers to know forwarding information. Congestion at any bottleneck will trigger consumers to reduce the requesting rate, thus the bandwidth utilization heavily relies on the performance of adaptive forwarding which aims to prevent overloading certain bottlenecks earlier than others. However, as adaptive forwarding tries to solve a distributed multi-commodity flow problem, it brings significant computational burdens to routers and lacks system scalability. Although some approximate solutions have been proposed, neglected factors (e.g. the path characteristics in OMCC-RF) cause noticeable performance degeneration (e.g. decreasing bandwidth utilization).

In fact, the performance problem will no longer exist if consumers control the amount of traffic on each bottleneck independently. To this end, this paper proposes PTP which lets consumers specify the adaptive forwarding (transmission path) of each Interest packet instead of routers. It is worth noting that, PTP does **NOT** entirely deprecate adaptive forwarding at the routers but uses it to probe the eligible transmission path. In addition, the routers' adaptive forwarding is also enabled when the specified transmission path of an Interest packet is invalid. The details are further presented in the following section.

### 3. Architecture Overview

In concept, PTP enables consumers to discover the eligible transmission paths and control the traffic on each one independently, which requires the support of both routers and consumers. Two main stages are involved in PTP, namely path probing and path-specified forwarding, where path probing corresponds to discovering eligible transmission paths and path-specified forwarding corresponds to traffic control.

*A. Probing Stage*

Initially, a new consumer does not have any prior knowledge of any transmission path in the network. Thus, the consumer needs to obtain some forwarding information from the network. In particular, a **tag** (as a part of an NDN packet) is introduced in PTP to record an eligible transmission path, where each unique tag corresponds to a unique path. In addition to notifying consumers of the transmission path, a tag can be appended to Interest packets, which enables routers to forward these Interest packets along the specified path. The detailed design will be discussed in Section 3-B.

In order to get the tag, the consumer periodically sends path probing Interest packets to the network. During the forwarding procedure of a probing Interest packet, the tag remains empty to prevent unnecessary overheads. When this Interest packet arrives at a producer, an empty tag is appended to the probing Data packet that corresponds to the probing Interest packet. This probing Data packet will be returned to the consumer along the Interest's transmission path but in the reverse direction. The tag is appended with the identifier of the receiving interface when the probing Data packet arrives at a router. Finally, the tag is returned to the consumer, and the consumer sets up an independent congestion control module for each tag that corresponds to an eligible transmission path. The encoding method is presented in Section 4-A.

Note that the probing Interest packets always bring back both tags and Data packets, which reduces the startup delays for probing paths. As the returned Data packets carry content that costs network resources, the frequency of probing is configured to a low value, e.g. 10 probing packets every second. In addition, MPFP creates a Forwarding Acceleration Base (FAB) for FIB lookup to bypass the costly LPM when forwarding Interest packets.

*B. Path-specified Stage*

Once a tag is obtained by a consumer, the consumer inserts the tag in the Interest packets that are waiting to be forwarded. Based on the tag in Interest packets, the router forwards the received Interest packets to the interface as specified, as shown in Figure 3.

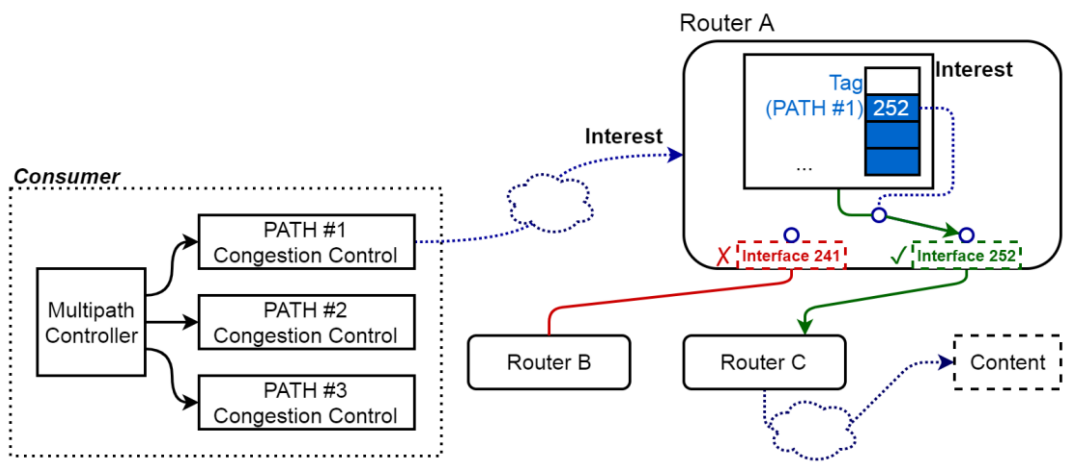

Figure 3  Path-specified Forwarding Engine

This formulates the path-specified forwarding, where the transmission path is controlled by consumers instead of routers. By adjusting the number of Interest packets sent to each specified path, the path-specified transport (e.g. congestion control) is formulated. To improve the efficiency, flexibility and robustness, one consumer is able to download content from multiple paths concurrently. However, PTP forces each consumer to keep at most *N* paths (configurable) that can be used concurrently, which prevents overusing network resources. For each transmission path, PTP uses a modified MPTCP congestion control scheme to guarantee multi-user/multi-flow fairness [6].

## 4. Multipath Path-specified Forwarding Protocol

This section will present the working principle and the detailed implementation of MPFP. It includes how the consumers collect forwarding information of eligible paths during the path probing stage, and how routers forward Interest packets according to the specified path during the path-specified forwarding stage.

*A. Path Probing and Tagging*

During the probing stage, when a router receives a probing Interest without a tag, the packet is forwarded to an interface according to an equal-weight strategy [6], which distributes packets to different interfaces evenly. It is worth noting that, the probing Interest packet is always forwarded to a producer, while the cache is ignored. The rationale is that the Data packets cached in routers may be unpredictable, as they may be affected by the caching policy. Thus, it is not a stable solution to obtain the path to a cache and control the traffic flowing from it. While the policies that cache continuous content at specific nodes (e.g. edge devices [12]) are exceptions. In this case, as the traffic flowing from these caching devices to the consumer is relatively stable, the consumer can treat the caching devices as a stable producer.

The consensual tag used in PTP is a stack, where each item is the identifier of a forwarding interface. The data structure of a tag is shown in Figure 4.

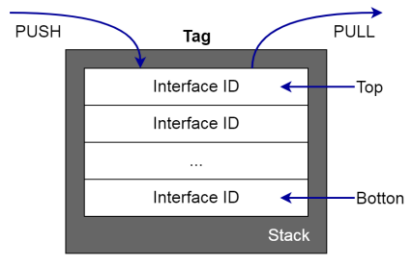

Figure 4 Data Structure of a Tag

As shown in Figure 5 (a), during the probing stage, an empty tag is appended to a Data packet by the producer. The (receiving) interface identifier of each hop is pushed into the tag until the Data packet is received by the consumer. As shown in Figure 5 (b), during the path-specified stage, a tag is appended to an Interest packet by the consumer. An item (interface identifier) is popped from the tag if the tagged Interest packet is received by a router. The popped identifier guides the router to forward the Interest packet to the specific interface. One advantage of this approach lies in its simplicity and its potential compatibility with the existing Multiprotocol Label Switching (MPLS) hardware.

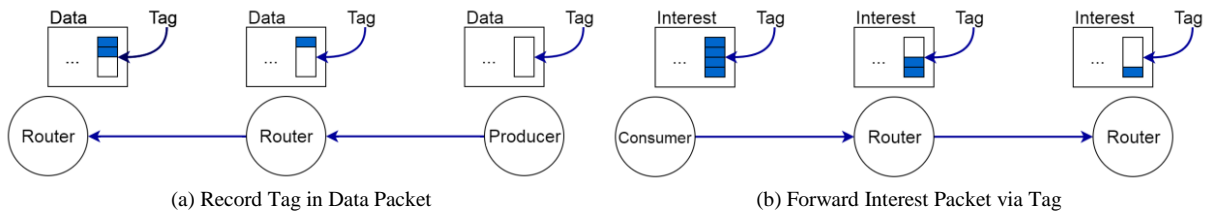

(a) Record Tag in Data Packet  (b) Forward Interest Packet via Tag

Figure 5 Examples of the Tagged Packets

In addition to stacking the interface identifiers, there are some other approaches that may be more efficient in terms of storage to construct the tag, such as using a data compression algorithm to encode these interface identifiers, which consequently reduce the bandwidth overheads. For example, it is possible to analyse the appearance frequencies of different interface identifiers over time, create the Huffman tree and compress the interface identifiers using Huffman coding. However, these advanced coding techniques may not be of significant benefit as the number of hops in an NDN network is relatively small. The overheads that can be saved are very limited. Moreover, the encoding/decoding and the analysing procedures can bring additional computational overheads to routers. The comparative study of applying different compression methods to optimize the performance is considered as future work.

### B. Forwarding Verification using Forwarding Acceleration Base

According to the popped interface identifier from the tag, the Interest packet is expected to be forwarded to the specified interface. However, the specific interface is not always valid, e.g. the next-hop device is disconnected or the connected producer no longer provides content. Therefore, the router must check if the specified interface is working before forwarding the Interest packet. A straightforward method is to first perform FIB lookup [1] using LPM, and then check if the specified interface exists in this FIB entry. As LPM is criticized for its expensive lookup overhead, this paper proposes a Forwarding Acceleration Base (FAB) to bypass LPM.

FAB aims to optimize the lookup of the elephant flows, where a flow is composed of a large number of packets with a shared prefix. For example, the name of a packet that belongs to an elephant flow is composed of a shared prefix (black) and a postfix (red) as shown in Figure 6. The longest shared prefix of the named packets is defined as the flow name, which can be obtained by removing the packet's postfix (i.e. the sequence number).

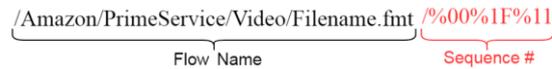

Figure 6 Name of a Data packet that belongs to a flow

The design principle of FAB is to create the low-complexity one-to-one mapping from the flow name to the FIB entry to replace LPM. FAB is the hybrid data structure of a hash table and a First-In-First-Out (FIFO) list. The hash table enables fast FIB lookup by flow names. The FIFO list records the flow name insertion order in the hash table to enable a FIFO replacement policy. The follow sections start from a simplified version of FAB (i.e. a hash table with unlimited size), and then proceed to the full version of FAB (i.e. a hash table with a limited size through a FIFO replacement policy).

*1) FAB: Hash Table without FIFO List*

The main part of FAB is a hash table which is an efficient data structure that stores the {*key*, *value*} tuples and it enables searching the *value* according to the *key* with the average-case time complexity of $O(1)$. The FAB records the flow name as the *key* and the FIB entry that corresponds to the flow name as the *value*. In other words, each tuple corresponds to the FIB lookup of a flow. An example of hash table is shown in Figure 7.

| Hashed Key | Tuple | |
|---|---|---|
| | Key [Flow Name] | Value [FIB Entry] |
| 0x01 | /apple/part1 | <ptr>fib-entry:{/apple} |
| 0x02 | /amazon/BBB | <ptr>fib-entry:{/amazon} |
| | ... | ... |
| | /AIT/sri/mvc_01 | <ptr>fib-entry:{/AIT/sri} |
| ... | ... | ... |
| 0x... | /UCLA/video/p1 | /UCLA |

Figure 7 Simplified FAB: Hash table without FIFO list

Here, the *value* is a pointer to the FIB-entry, i.e. *<ptr>fib-entry*, the *key* is the flow name, and the *hashed key* is the result of calculating the hash value of the *key*. To avoid hash collisions, where multiple *keys* are mapped to a single *hashed key*, each *<key, value>* tuple that is mapped to the single hashed key is recorded in the table. The FAB lookup process is as follows. Firstly, the input flow name (e.g. /amazon/BBB) is converted into the hashed key (e.g. 0x02). Then all the tuples that correspond to this hashed key are obtained. If the input flow name does match a key in these tuples, the correspondent FIB entry is returned. Note that the FIB-entry already contains the flow prefix (as a sub-string of the flow name), thus FAB can compare the input flow name and the flow prefix (in the FIB-entry) without storing the *key*. This design can further reduce storage overheads.

If the range of the hashed key is set to a large value, the number of tuples that corresponds to each hashed *key* will be sufficiently small (i.e. approximately equal to 1). This implies that the lookup cost of FIB lookup using this hash table is much lower than LPM.

Note that when a router is initialized, the hash table is empty. The insertions of tuples are triggered by the arriving Interest packets. Particularly, when the router receives an Interest packet, the flow name of the Interest packet is searched inside the hash table. If this flow name does exist as a *key*, the router forwards the Interest packet according to the matched FIB entry. Otherwise, the router applies LPM to the flow name to get the FIB entry. Then, the flow name and the pointer to the FIB entry are inserted into the hash table for the later lookup.

2) *FAB: Hash Table + FIFO List*

The number of tuples in the hash table grows along with the increasing number of flows. To prevent wasting storage, the number of tuples in the hash table is limited to a suitable value. According to the previous study [13], the content popularity in content-centric networks follows the Zipf-Mandelbrot law. This is to say, the majority of Interest packets in NDN belong to few active flows. Under this assumption, maintaining the hash tables only for active flows reduces the lookup overheads for most Interest packets and does not cost much storage resources. A light-weight strategy – FIFO can be employed to retain the active flows in and evict the inactive flows from the hash table. However, the vanilla hash table lacks the ability to insert and to evict tuples according to the FIFO policy. To this end, an auxiliary data structure – FIFO list is integrated with the hash table. An example is shown in Figure 8.

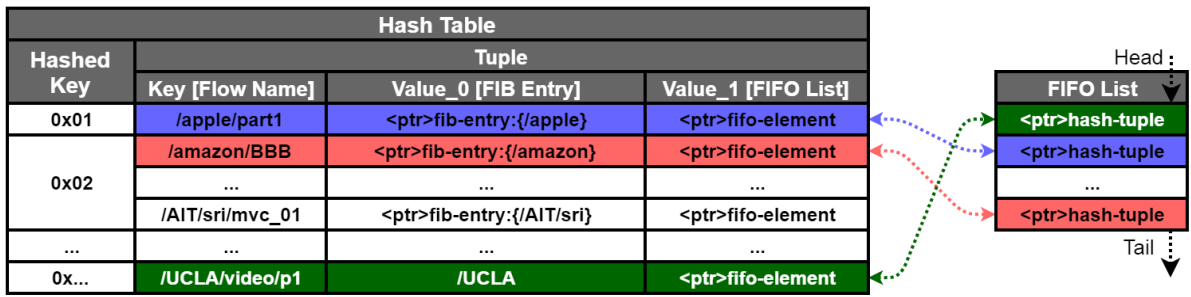

Figure 8 FAB: Hash table + FIFO List

Here, <ptr>hash-tuple denotes the pointer to a tuple in the hash table. The intuition of this hybrid data structure is to record the inserting order of each tuple in the hash table using a FIFO list (i.e. doubly linked list), such that the tuples can be inserted and evicted according to the FIFO strategy. This is implemented by interlinking (dash lines in Figure 8) the tuple in the hash table and the element in the FIFO list. Particularly, when a new tuple is inserted into the hash table, a new element is also added to the head of FIFO list. The element in the list is a pointer to the tuple and the *value*_1 in the tuple is a pointer to the element in the list. Based on this design, when the hash table is full and it needs to evict the earliest tuple, it locates the earliest tuple according to the tail element of the FIFO list, and it deletes the tail element in the list and also the correspondent tuple. In addition, if a flow name of an incoming Interest packet matches the key of a tuple, the tuple should be treated the tuple as a newly inserted tuple. This is realized by locating the element in the list via the *value*_1 in the tuple and shifting the element to the head of the list.

As the Interest packets that belong to an active flow frequently update the tuple by shifting the linked list element to the head of the list, active flows have larger chances to stay in the head of the list. This indicates the linked tuples of the active flows will be retained in the hash table. Thus, the Interest packets that belong to the active flows (i.e. the majority of the Interest packets) can use FAB to access the FIB entry to reduce the lookup overhead.

   *3) Remark*

It is worth to note that the eviction of the tuples does not affect the forwarding decision of any flow. The Interest packets that belong to the inactive flows (i.e. not in FAB) can always use LPM to access the FIB entry. In addition, FAB is not limited to MPFP. It can be integrated into the original NDN forwarder daemon to reduce the lookup cost. However, this is beyond the scope of this paper and is considered as a future work.

*C. Transmission Path Selection and Management*

Via probing, a consumer may capture a number of eligible paths for transmission according to the received tags. To control the management complexity [2] or to satisfy the QoE requirements (e.g. latency) of the application, the consumer may need to choose a subset of concurrent paths for transmission (in-use path), while marking the remaining paths (unused paths) as disabled. Optimisation of path selection is beyond the scope of this paper. In MPFP, a bandwidth-based selection strategy is employed to choose the in-use paths for illustration purpose. Some other feasible selection strategies are further discussed in Appendix I.

   *1) Bandwidth-based Selection*

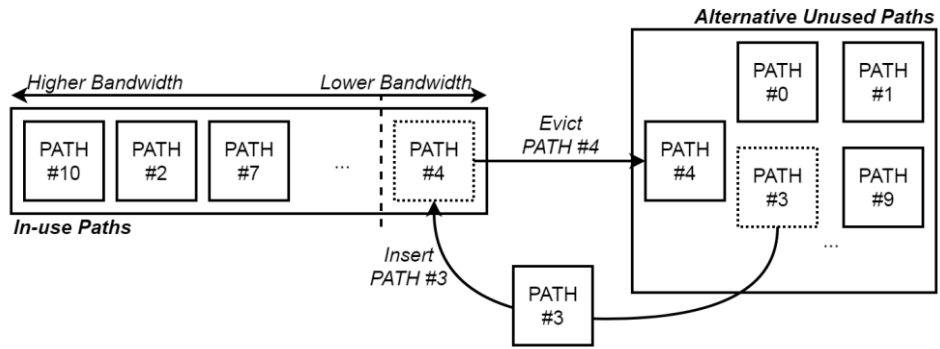

Figure 9 High Bandwidth Selection Strategy

As PTP is motivated by improving downloading bandwidths, the consumer needs to select the paths with higher bandwidths. As the bandwidth information can be hardly perceived from the probing packets[1], a greedy algorithm is introduced here to continuously select the paths. The greedy algorithm demands the consumer to periodically sort the bandwidths of its *N* in-use paths in descendant order. The in-use path with the lowest bandwidth is replaced by an unused path.

---

[1] This paper does not consider the packet pair or the packet train approach to measure the bottleneck bandwidth.

An example is shown in Figure 9. The consumer sorts the bandwidth for *N* paths and the result is PATH #10, PATH #2, PATH #7 … and PATH #4. As PATH #4 has the lowest bandwidth, it is replaced by a random unused path, e.g. PATH #3. After a certain time period *T* (configurable), the consumer re-sorts the bandwidths for the in-use paths.

*2) Other Consideration*

In some cases, such as a link (a hop) in an in-use path fails, the router will proactively return NACK packets to notify the failure of the path. Based on the received NACK packets, the consumer disables this failed in-use path and replaces it with one unused path. Additionally, to prevent consumers frequently replacing the in-use paths, each consumer is limited to change the in-use paths every *T* (configurable) seconds.

## 5. Multipath Path-specified Congestion Control Protocol

By specifying the transmission path for each packet, the congestion control of all in-use paths becomes very similar to that in MPTCP, where the consumer maximizes utilization, minimizes congestion and balances fairness. Differently, the routers in NDN can cache content temporarily, which enables consumers to retrieve Data packets from intermediate routers without accessing the original server. In this case, the Data packets from the intermediate nodes will affect the round-trip time, and consequently affect the detection of congestion, e.g. estimating the Retransmission Timeout (RTO). Multipath Path-specified Congestion Control Protocol (MPCCP) is a tailored transport protocol for PTP, which enables consumers to distinguish the traffic from the original server and that from the intermediate nodes. Note that the MPCCP of the current version is to verify the feasibility of PTP but is not targeting specific performance requirements. For simplicity, an off-the-shelf MPTCP control law is employed. Additionally, a novel congestion detection scheme is proposed in MPCCP to fit the NDN's caching/aggregation characteristics.

*A. Similarity and Different to MPTCP*

Without the considerations of in-network cache and traffic aggregation, the traffic pattern in PTP is very similar to that in MPTCP. The trivial difference is that MPTCP is sender-driven but NDN is receiver-driven, which does affect the operating logic of a congestion control law. In this respect, a loss-based MPTCP control law, Linked Increase (Semi-coupled[2]), is directly ported to NDN to deal with the fairness issues [6].

However, the congestion detection logic used in TCP is no longer valid in NDN. This is because the Data packets may be returned from intermediate routers (cache or traffic aggregation). The packets that come from the intermediate nodes usually lead to smaller round-trip times and may arrive ahead of the previously requested packets. A consumer cannot predict if the Data packets are from cache/aggregation or not. Thus the consumer has to assume the worst situation by default, which is that all packets are from the producer. In principle, the congestion that happens on a sub-section of a full in-use path is also considered as the congestion of the in-use path.

*B. Assumption*

***Fixed Size Data Packet:***

---

[2] Semi-coupled: it tries to maintain a moderate amount of traffic on each path while having a bias in favor of less congested paths

Each content object that belongs to the same flow is divided into fixed-size packets, where the number of packets *N* is negotiated by the producers and the consumers [14] in advance. The sequence number of a packet in the object is abbreviated to SEQ in this section. Note that a fixed packet size is a practical assumption for the elephant flows that deliver large content files or video chunks. In case of mice flow (data generated by sensors), the size of a Data packet may be unpredictable, and the name-based routing/forwarding is a more feasible solution.

*Congestion State per Full Path:*

Cache and aggregation cause the traffic to start from the intermediate nodes in the full path (i.e. from producer to consumer). The traffic within the subsection may also cause congestion. From the perspective of consistency, the congestion of any link within the full path is considered as the congestion of the full path, and the traffic on the full path is controlled as an entirety, as shown in Figure 10.

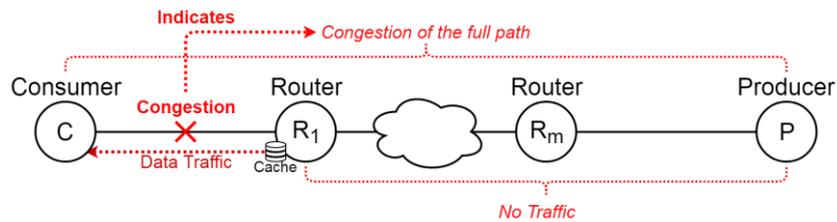

Figure 10 Congestion of a path and its sub-section

## C. Design Principle

### 1) Fast Loss Detection

Unlike TCP which detects loss by 3 duplicated ACKs, NDN uses Data packets for acknowledgement, and loss is detected by unreceived Data packets. As Interest packets are delivered to multiple in-use paths concurrently, the SEQ of Interest packets that are forwarded to an in-use path can be discontinuous. To this end, MPCCP lets consumers record the forwarding order of the Interest packets to each in-use path, and the corresponding Data packets (from producers) are expected to arrive in the same sequence. To realize an efficient and reliable loss detection scheme, consumers check the arriving sequence of each Data packet. If the Data packets are from intermediate routers (i.e. the transmission latencies of these packets are smaller than expectations), they are ignored for checking the arriving order and updating the RTO. In most cases, the round-trip time of travelling the full path is always longer than that of travelling its sub-sections. Under this assumption, once a Data packet that is returned from the producer (i.e. travels the full path) arrives, all the previous Data packets[3] are expected to be received. Otherwise, the loss is detected. Based on the principle of loss-based congestion control, the lost packets are re-transmitted, and the congestion window of this in-use path is decreased. In practice, the lost packets are prioritized for re-transmission on the same in-use path, as the lost packet may be cached somewhere along this in-use path.

### 2) Timeout Detection

In addition to detecting congestion based on the arrived sequence, consumers also employ timeouts for loss detection. There are two reasons for this. The first one is that all packets may be lost in highly congested links. As no packet is received, the consumer cannot detect congestion via checking the receiving orders of Data packets. The second one is that all packets are returned from cache/aggregation, where all Data packets are ignored for checking orders. For both cases, consumers use RTO timers (as in TCP) to check if packets are dropped. If a packet is timed out, the consumer will consider the network to be highly congested. Thus, the congestion window is set to the minimum value and is increased (multiplicatively or additively) until it reaches the slow-start threshold.

---

[3] Previous Data packets: the corresponding Interest packets were sent earlier

3) *Remarks*

Ultimately, the aim of MPCCP is to detect congestion based on the received sequence of the Data packets (that travelled the full path) and the timeout for each packet. Thus, the congestion window is adjusted according to a loss-based control law. A latency-based control law will be considered in future work.

D. *Implementation*

A tag is placed in each Interest packet to ensure path-specified forwarding. As Data packets are routed via PIT, there is no need to place tags in Data packets. In order to distinguish which path a Data packet comes from, the consumer records the forwarding in-use path in a vector indexed by SEQ. When a Data packet arrives, the consumer locates the $SEQ^{th}$ (carried by the Data packet) element in the vector to get the forwarding in-use path of the Data packet. Then, it performs congestion detection and congestion control on this in-use path. Similar to TCP, there are 3 states for each in-use path: Congestion Avoidance, Fast Recovery and Slow Start. The state transfer graph is shown in Figure 11.

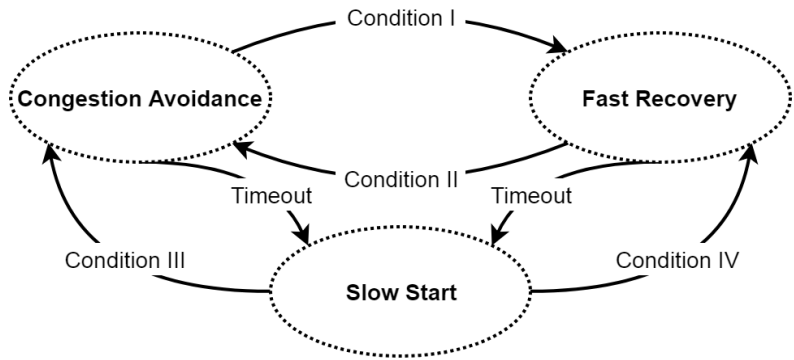

Figure 11 State Transfer Graph for the proposed Congestion Control

1) *Congestion Avoidance*

In order to detect packet loss, the consumer maintains a queue for each in-use path to record the forwarding sequence of each Interest packet (the queue is indexed by the forwarding sequence), and each element (in the queue) records the SEQ and the receiving status (RCV) of this packet. If a Data packet is received, the window of the in-use path is increased additively and the element in the queue is located using the SEQ (carried by the Data packet). The RCV of this element is set to 1. If this Data packet is returned from an intermediate router (the router labels the Data packet using a 1-bit flag), the consumer ignores this packet for congestion detection of the corresponding in-use path. If the Data packet is from the producer, the consumer updates the in-use path RTO and then checks if all previous Data packets are received from the in-use path. **Condition I**: If the RCV of any previous element is 0 (not received), the packet loss is detected. The lost packets are recorded in another queue for re-transmission. Moreover, the congestion window is decreased multiplicatively. Finally, the in-use path enters fast recovery state.

2) *Fast Recovery*

If an in-use path enters the fast recovery state, the consumer memorizes the current congestion window size, $cwnd_{bk}$. For each received Data packet, the congestion window is increased additively, which is identical to that in the congestion avoidance state. If a loss is detected again, the size of congestion windows is set to $cwnd_{bk}$. This is because loss indicates that the congestion in the network has not finished yet. Furthermore, if any loss happens after the in-use path has received $cwnd_{bk}$ Data packets, the congestion window is further decreased by multiplicative decrease again and reenters the fast recovery state. **Condition II**: If a backup window size of Data packets is received continuously without losses, the sub path quits to the congestion avoidance state.

*3) Slow Start*

As in TCP, the initial state of an in-use path is slow start, which enables probing of the available bandwidth. During this stage, the congestion window is increased multiplicatively if a Data packet is returned from a producer, and the congestion window is increased additively if a Data packet is returned from an intermediate node. This prevents overshoot of the congestion window if the current Data packets are all returned from cache/aggregation (i.e. with larger bandwidth) but the following packets are from the producer (i.e. with smaller bandwidth). **Condition III**: the in-use path will quit the slow start state and enters the congestion avoidance state if the congestion window reaches the slow-start threshold. Moreover, **Condition IV**: if any packet loss is detected, the in-use path will enter the fast recovery state directly. On the other hand, if a packet is timed out, the threshold is cut in half and the in-use path enters slow start, no matter what the current state of the in-use path is.

*4) Inflight Control*

In PTP, the calculation of inflight is different from that in TCP. As TCP cannot determine how many packets are staying in the network based on cumulative ACKs. When a duplicate ACK is received, the sender increases the congestion window but does not decrease the inflight packet (because the TCP receiver may drop the out-of-sequence packets when memory is full). In PTP, the lost and acknowledged Data packets trigger a reduction in the number of inflight packets as the consumer stores the out of sequence packets. The consumer can drop these packets and re-send corresponding interests if required.

*5) Control Law*

Designing an optimal control law is not the focus of this paper. In general, all MPTCP control laws can be used in MPCCP (e.g. EWTCP [15], Coupled [6] and wVegas [16]). This paper evaluates performance using a mature approach – Linked Increase (Semi-Coupled) for additive increase and multiplicative decrease. The additive increase is given in *eq.* (2)

$$cwnd_p \Leftarrow cwnd_p + \frac{rtt_p}{rtt_\mu} \min\left(\frac{\alpha}{\sum_{\pi \in \mathbf{P}} cwnd_\pi}, \frac{1}{cwnd_p}\right) \quad (2)$$

where $cwnd_p$ denotes the congestion window of path $p$, $\sum_\pi cwnd_\pi$ denotes the sum of the congestion windows of all in-use paths $\pi \in \mathbf{P}$ and $\alpha$ is a parameter that controls the aggressiveness to achieve sub-flow fairness [6]. $rtt_\mu$ denotes the reference RTT [17] and $rtt_p$ denotes the path RTT. $rtt_\mu / rtt_p$ compensates the RTT unfairness that makes longer RTT flows obtain less bandwidth. The parameter $\alpha$ is calculated automatically in Linked Increase, as shown in *eq.* (3)

$$\alpha = cwnd_{total} \frac{\max_{\pi \in \mathbf{P}} \left( \frac{cwnd_\pi}{rtt_\pi^2} \right)}{\left( \sum_{\pi \in \mathbf{P}} \frac{cwnd_\pi}{rtt_\pi} \right)^2} \quad (3)$$

$$cwnd_{total} = \sum_{\pi \in \mathbf{P}} cwnd_\pi$$

where $rtt_\pi$ denotes the round-trip time of path $\pi$. Linked Increase reduces the per-path congestion window as in TCP, which multiplies the per-path congestion window $cwnd_p$ by a positive constant value $\beta$ (configurable and is smaller than 1).

$$cwnd_p \Leftarrow \beta \times cwnd_p \quad (4)$$

*E. Other Considerations*

**Fast Loss Detection:** In this paper, we assume the packet loss is detected by one reordered Data packet from the producer. This can be inaccurate if network devices proactively reorder the sequences of the packets. An extended scheme is to detect the loss based on two Data packets from the producer. In particular, once two Data packets (from the producer) arrive, the Data packets before the first one should have been all received; otherwise, congestion is detected.

## 6. Evaluation

This section demonstrates the correctness and effectiveness of PTP. The performance is evaluated from three perspectives: 1) bandwidth utilization by multiple sources/paths, 2) fairness among users and 3) reaction time to in-network caching.

For comparative studies, the state-of-the-art approach – OMCC-RF is selected, as it provides both the congestion control scheme (RAAQM) and the adaptive forwarding strategy (RFA). Additionally, the cross combinations of the congestion control schemes (CHoPCoP [7] and RAAQM) and the forwarding strategies (PAF [3] and RFA) are used for evaluations. The rationale of stitching these approaches is that many existing congestion control schemes (CHoPCoP, ECP [18]) do not consider the multipath forwarding and the adaptive forwarding strategies (PAF [3], SAF [19], EPF [20]) lack the discussions on multipath congestion control, which makes them hard to be compared with PTP.

The approaches to be evaluated in this section are listed below.

*Sol.1*: Path-specified Transport Protocol (PTP) (MPFP + MPCCP), proposed by this paper

Path-specified Transport Protocol (PTP) is an approach that let consumers control the traffic on each transmission path independently. It includes two sub-protocols – Multipath Path-specified Forwarding Protocol (MPFP) and Multipath Path-specified Congestion Control Protocol (MPCCP).

*Sol.2*: Optimal Multipath Congestion Control and Request Forwarding (OMCC-RF) [2] (RFA + RAAQM)

OMCC-RF aims to optimize a Network Utility Maximization problem in a distributed manner. Particularly, the problem is solved by performing congestion control (i.e. RAAQM) at end nodes and adaptive forwarding (i.e. RFA) at intermediate routers.

*Sol.3*: RFA [2] + CHoPCoP [7]:

As discussed in the Section 2.C, OMCC-RF suffers from an issue that the forwarding strategy (RFA) requires a large queuing length to improve throughput, while the congestion control (RAAQM) operates at a smaller queuing length (i.e. it is a latency-based scheme). To fix this contradiction, the RFA strategy is combined with an ECN-based (queue-based) congestion control scheme – Chunk-switched Hop Pull Control Protocol (CHoPCoP) which enables the bottleneck to queue sufficient packets. Additionally, CHoPCoP employs a hop-by-hop flow-aware shaping for intermediate routers.

*Sol.4*: PAF [3] + RAAQM [7]:

In addition to the PI-based forwarding strategy, the delay-based strategy is also worthy to be compared. PAF is inspired by the ant colony optimization algorithm (ACO) to balance the load on different paths based on RTT. This stitching approach is to illustrate that the latency-based strategy also requires long queues to guarantee throughput. When it is cooperating with the delay-based congestion control (e.g. RAAQM), the performance is poor.

A. *Simulation and Parameter Settings*

PTP is implemented through the NDN Forwarding Daemon and evaluated via ndnSIM [21]. The payload size is set to 1024 bytes. For the configurable parameters, the number of in-use paths $N$ is set to 10; the time interval $T$ which lets consumer switch the in-use paths is set to 10s, and the decreasing factor $\beta$ is set to 0.75.

B. *Scenario 1: Tree Topology*

In the first experiment, a consumer (#1) is downloading content from 4 producers (#5, #6, #7, and #8). The topology is a tree structure as shown in Figure 12.

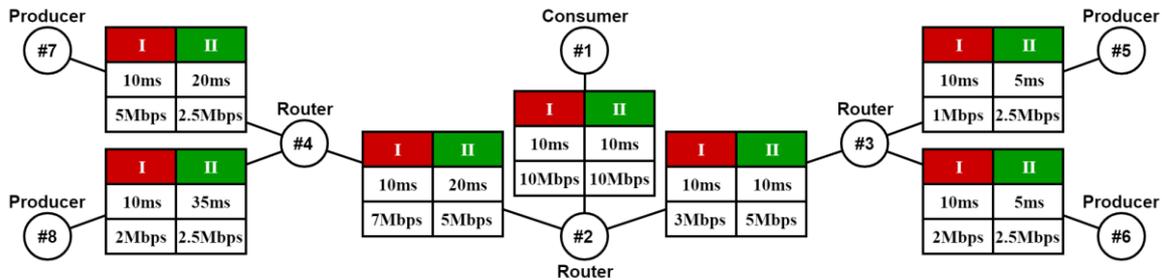

Figure 12 Tree Network Topology

Two types of link settings are considered. In Case I, the latencies of different transmission paths are identical, but the bandwidths are different values. This setting is to test if different solutions can fully utilize the bandwidth of all paths. In Case II, the bandwidths of different transmission paths identical, but the latencies are different. This experiment tests if the transport protocol has a bias on latency.

**Comments on the results:** Figure 13 and Figure 14 report the averaged utilization and the throughput of each link, where the utilization is calculated by dividing the measured throughput by the theoretical throughput (calculated by max flow) for each link. As consumer #1 is connected to the content via a unique link 1-2, where all traffic is aggregated in this link, the effective downloading rate is equal to the throughput of link 1-2. In Case I, the link utilization of Sol.1 (PTP) is nearly 97.78% and that of Sol.2, Sol.3 and Sol.4 are around 81.56%, 86.89%, and 79.11% respectively. All solutions can shift congestion from the congested paths to the congestion-free paths. However, as the congestion metric – PI (i.e. used in Sol.2 and Sol.3) is not accurate enough to reflect the available bandwidth (i.e. as discussed in Section 2.C), they will overload the paths with lower bandwidth first.

As Sol.3 uses an ECN-based congestion detection, the backlogging queue helps in compensating the PI metric, which consequently improves throughput and utilization. In Sol.4, as RTT reflects the congestion levels of different links, it achieves the similar performance as Sol.3.

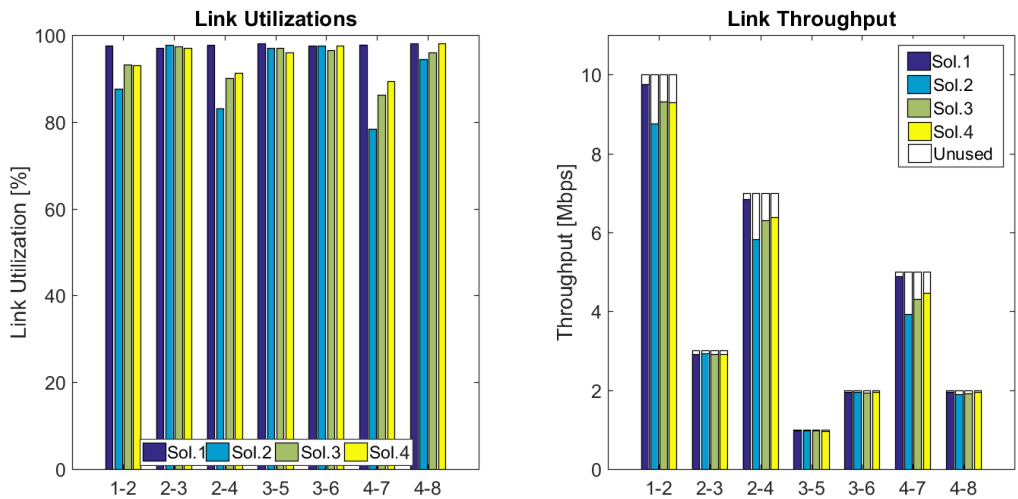

Figure 13 [Scenario 1] Case I Utilization/Throughput of each link

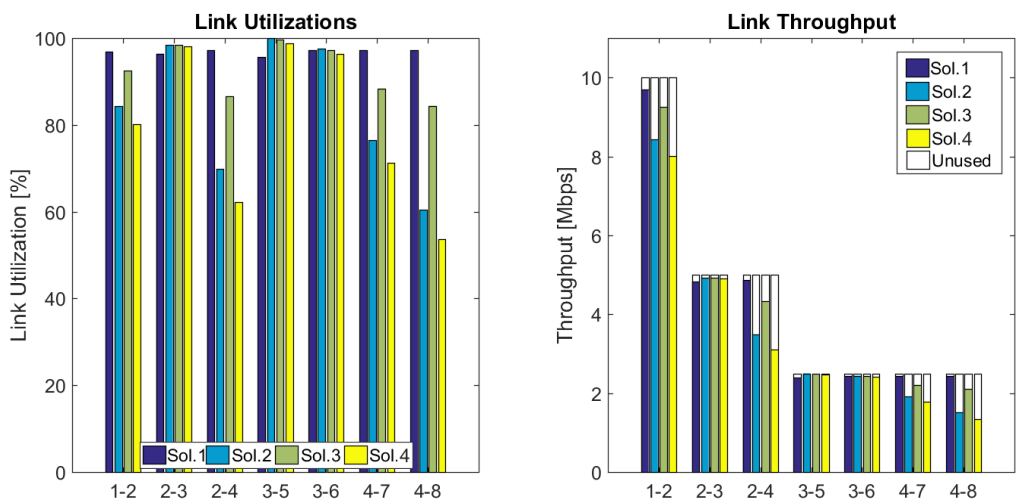

Figure 14 [Scenario 1], Case II Utilization/Throughput of each link

In Case II, the link utilization of Sol.1 (PTP) is nearly 96.9%, and that of Sol.2, Sol.3 and Sol.4 are 84.3%, 92.5% and 80.1% respectively. As the solutions except PTP are based on the latency metrics (i.e. PI and RTT), they suffer from latency bias [4], they always overuse the low latency paths. Specifically, the path 1-2-3-5 and the path 1-2-3-6 are with the smallest latency (25ms), the bandwidths of these paths are fully utilized. As these two paths reach congestion earlier than the other paths, RAAQM (in Sol.2 and Sol.3) decreases the congestion window to all paths as they are congested (i.e. RTT increases). This results in insufficient bandwidth utilization. For Sol.4, as CHoPCoP is configured to queue sufficient packets in the bottleneck (3-5 and 3-6) to compensate PI, the throughputs of the other paths are increased. However, the large queue settings lead to unexpected RTO timeout and returning a large number of ECNs. Based on these congestion signals, the consumer decreases the requesting rate thus degrades the bandwidth utilizations. In contrast, as PTP controls the sub-flow on each path independently, it does not suffer from the problem encountered by the conventional framework. The result shows that PTP achieves the best bandwidth utilizations compared to the other solutions.

C. *Scenario 2: Diamond Topology*

The second experiment aims to verify the effectiveness of PTP in a diamond topology [22], which can be used to test the performance of the cases with shared links. The traffic is aggregated and split at certain nodes. Different from the tree topology which split traffic to multiple paths, the diamond topology also allows traffic from different paths to be aggregated. In the real world topology, the routing paths do not always follow the tree structure. This topology is used to verify the performance of more general multipath situations. In this experiment, a consumer (#1) is connected to the content object via 4 joint paths (Figure 15, 1-2-3-4-6-7, 1-2-3-5-6-7, 1-2-4-6-7 and 1-2-5-6-7).

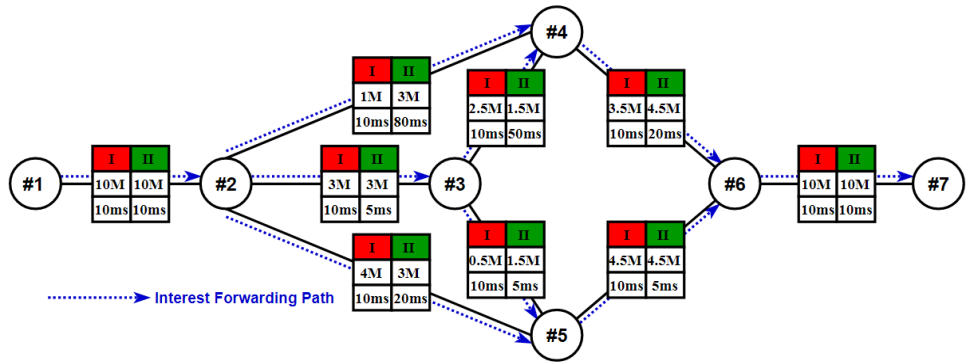

Figure 15 Diamond Network Topology

Following the same consideration as that in the tree topology (Figure 15), two different link settings are considered, with the same latencies and different bandwidths (Case I) and with the same bandwidths and different latencies (Case II). Note that the bottlenecks are the links within the diamond (#2, #3, #4, #5, #6, calculated by max-flow).

**Comments on the results:** Figure 16 and Figure 17 report the averaged link utilization and throughput in both cases in comparison with the optimal usage. Obviously, the existing solutions (Sol.2, Sol.3 and Sol.4) have difficulty in supporting this topology in both cases, and they achieve worse performances in Case I. In Case I, the link utilization of Sol.1 (PTP) is nearly 97.1%, and that of Sol.2, Sol.3 and Sol.4 are 41.5%, 64.6% and 35.6% respectively. As the bandwidth differences are relatively large, congestion always happens at the bottlenecks with much lower bandwidths (3-5 and 2-4). Although path 2-5-6-7 does not suffer from congestion, and node #2 tries to forward to node #5 with more packets. However, as the consumer (#1) detects the congestions at the bottlenecks (3-5 and 2-4) timely and reduces the congestion window immediately. This limits the utilization of link 2-5-6-7 to a low level.

In Case II, the link utilization of Sol.1 (PTP) is 97.3%, and that of Sol.2, Sol.3 and Sol.4 are 81.1%, 85.2% and 78.6% respectively. As the latency of path 1-2-3-5-6 is relatively smaller than others (the numbers of PIs at #2 and #3 are less), it is congested earlier and it triggers the congestion control at consumers. In general, Case II achieves better utilizations than Case I. This can be explained as the measurements of PI and RTT are averaged by the sharing paths. This makes the forwarding strategy more evenly distribute traffic on different paths. As the bandwidth of different paths is similar in Case II, the bandwidth utilization in Case II is better than that in Case I. As the congestion control in PTP follows the Linked Increase algorithm, it can smartly shift the traffic from the more congested paths to the less congested paths to incrementally increase the bandwidth utilization. PTP outperforms existing solutions and achieves near-perfect utilizations on all paths.

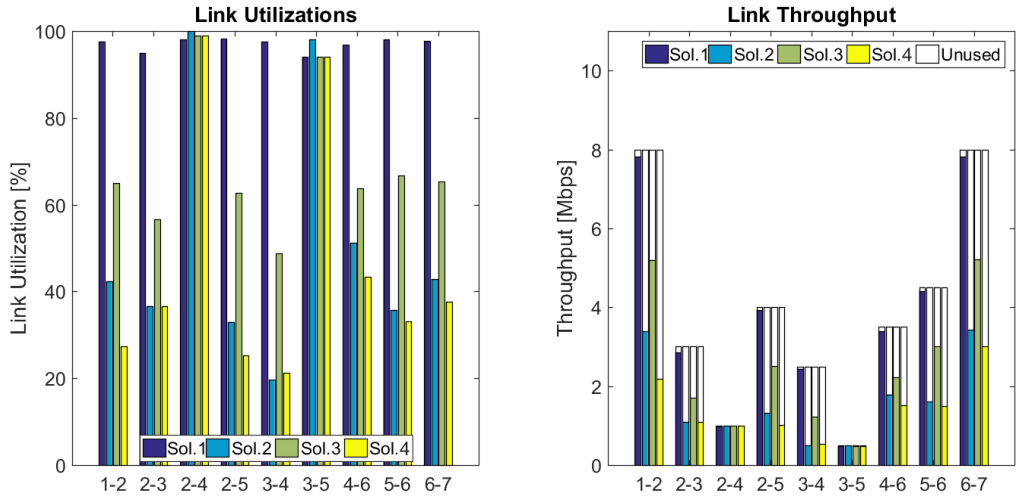

Figure 16 [Scenario 2] Case I Utilization/Throughput of each link

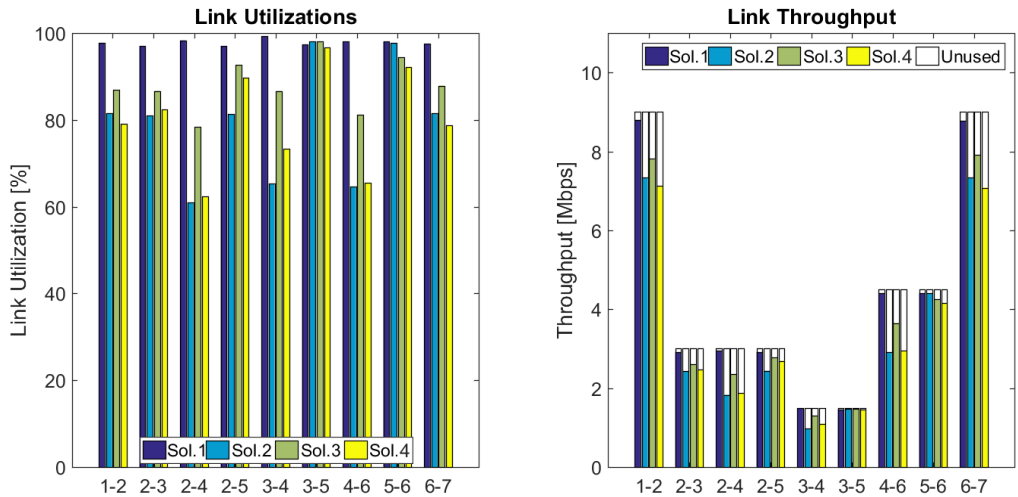

Figure 17 [Scenario 2] Case I Utilization/Throughput of each link

D. *Scenario 3: Multi-Flow vs Single-Flow*

A significant problem in MPTCP is fairness [6] when a single-flow and a multi-flow are competing for the same bottleneck. Although the MPTCP control law is employed, it is the first time that the control law is used in a receiver-driven model. This experiment verifies the feasibility of this control law using the topology shown in Figure 18. Consumer #1 is downloading Data packets of the content object **A** from Producer #5 and Producer #6, and Consumer #2 is downloading Data packets of the content object **B** from Producer #5 only. The shared bottleneck is 3-4.

**Comments on the results:** Figure 19 and Figure 20 illustrate the variations of congestion window sizes of Sol.1 (PTP) and Sol.2 (OMCC-RF). Obviously, the fairness performance of PTP is much better than that of OMCC-RF. As shown in Figure 19, PTP achieves the fairness ratio of 48:52. This is because PTP is based on Linked

Increase (Semi-coupled) that inherently supports multi-flow fairness. Figure 20 shows that multi-flow in Sol.2 (OMCC-RF) is more aggressive than single flow, and the fairness ratio is 38:62. Similar to Sol.2, as Sol.4 uses RAAQM for congestion control, it achieves the fairness ratio which is 41:59. As Sol.3 employs the flow-aware hop-by-hop Interest shaping scheme, the bandwidth allocation to each consumer is almost the same. Its fairness ratio is 49:51. Although the hop-by-hop Interest shaping achieves remarkable inter-flow fairness by explicitly allocating bandwidth to each flow, two noticeable problems are brought by it. First, the flow-level fairness is not equivalent to user-level fairness. If the number of the consumers that belong to one flow is larger than that belong to another flow, equally dividing bandwidth to two flows do not always mean equally allocating bandwidth to consumers[4]. The second is that, as the network may have numerous flows between different nodes, letting routers balance the traffic for each flow lacks system scalability.

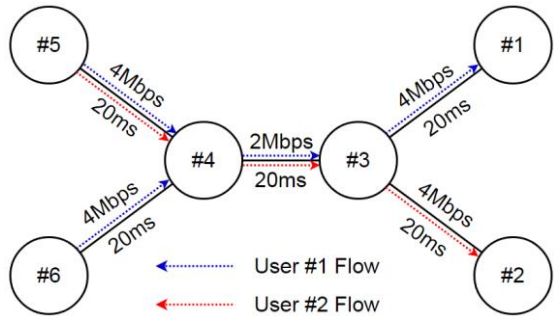

Figure 18 Topology for Multi-Flow vs Single-Flow

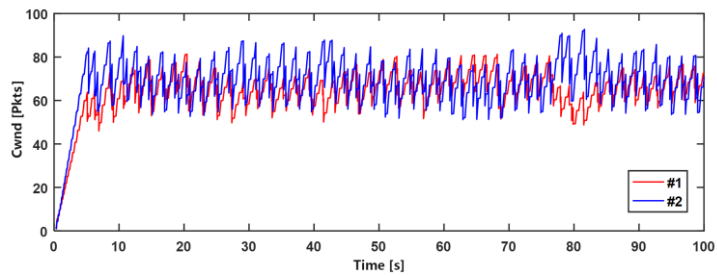

Figure 19 [Scenario 3] Fairness Convergence using PTP

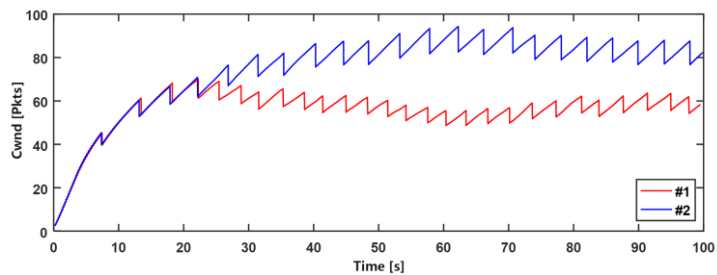

Figure 20 [Scenario 3] Fairness Convergence using OMCCC-RF

### E. Scenario 4: Cache Support

The fourth experiment verifies the PTP's ability to track the bandwidth provided by cache. The metric is the response time of two situations. They are cache appears and cache disappears. The topology is shown in Figure 21. In the first 20s, consumer #5 downloads content from producer #4. Note that consumer #5 only download 2000 Data packets from #4, i.e. after that consumer #5 stops requesting content. Half of the 2000 Data packets (randomly selected) are cached in node #3. After 20s, consumer #5 is stopped, and consumer #1 starts to download content from #4. The bottleneck between #1 and #5 is 1Mbps (5-3) if cache is not available. In the case that the content is fully cached in #3, the bottleneck between #1 and #3 is 2Mbps (3-2). As half of the first 2000 Data packets (from #4 to #5) are already cached in node #3, the ideal bottleneck should be around 2Mbps (3-2), unless the 2000 Data packets are downloaded. In other words, 1Mbps bandwidth can be provided by producer #4 and 1Mbps bandwidth is provided by the cache of 1000 packets in node #3.

---

[4] If the consumers are requesting the same content, the inter-user fairness is the same as the inter-flow fairness. Otherwise, the bandwidth received by a consumer is inversely proportional to the number of consumers that belong to the flow.

**Comments on the results:** As shown in Figure 23, the downloading rate of consumer #1 first increases to nearly 1.9Mbps to nearly fill the 2Mbps bottleneck (3-2) where 0.9Mbps is from the cache and the other 1Mbps is from producer #4. After the 1000 cached Data packets are downloaded, the usable bandwidth is dropped to 1Mbps (4-3). This experiment verifies that PTP can effectively support in-network cache, as the Data packets from cache are out-of-order compared to the packets from the producer, but the congestion control scheme still can near-perfectly utilize the bandwidth. Furthermore, Figure 22 shows that the consumer can fast track the availability of cache. When the cached content becomes unavailable, the Interest requesting rate is reduced to the suitable value, which lets the returning Data packets properly fill the 1Mbps bottleneck.

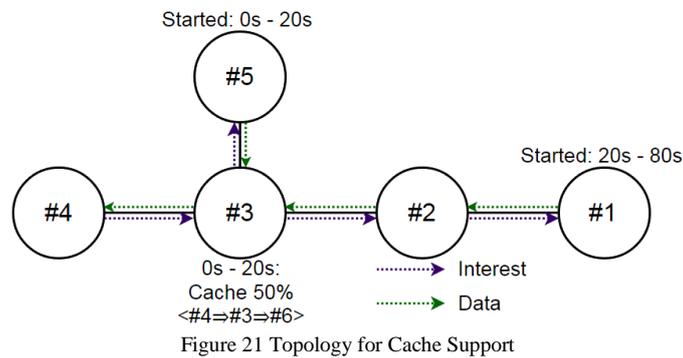

Figure 21 Topology for Cache Support

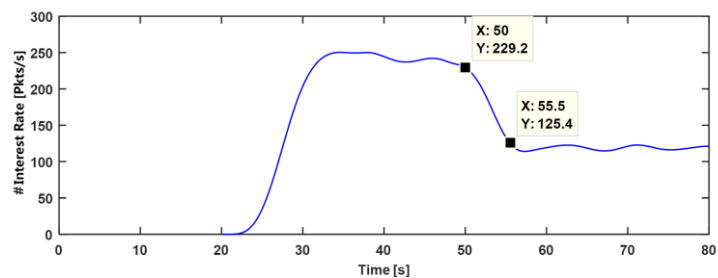

Figure 22 [Scenario 4] Interest request rate along time

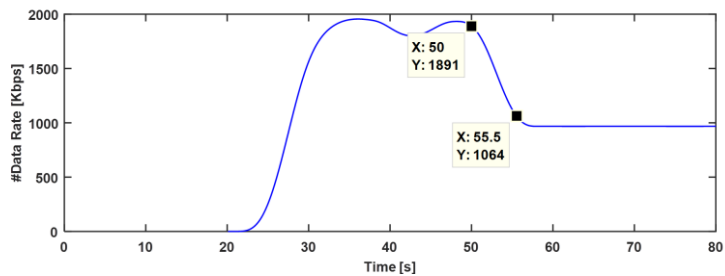

Figure 23 [Scenario 4] Data download rate along time

*F. Scenario 5: Hierarchical-Mesh Topology*

Consider a complex topology, as shown in Figure 24. The bandwidths of links 2-1, 3-2, 4-2, 8-2 are 1400Kbps, 400Kbps, 600Kbps and 400Kbps respectively, and the other link's bandwidths are 200Kbps. The latency of each link is set to 20ms. Due to the presence of redundant links (#3-#4, #5-#7), the throughput of the network may be reduced if the two paths are filled[5]. This experiment aims to verify if the congestion control scheme is able to avoid using the redundant links to maximize the utilization.

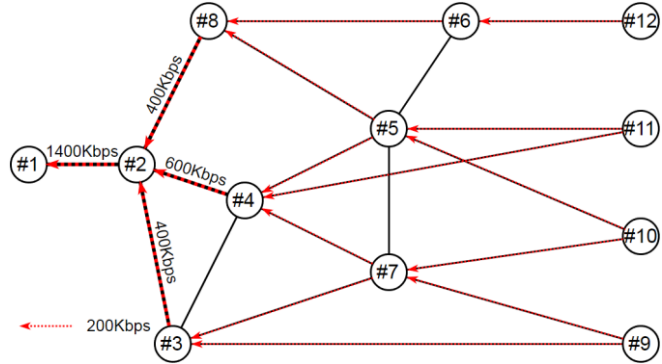

Figure 24 Max-flow support for a hierarchical topology

**Comments on the results:** In Figure 24, the dash lines denote the optimal in-use links that are calculated by solving a max flow problem (1400Kbps). Table I shows the actual utilization of each link based on PTP. The result shows that PTP can achieve very high bandwidth utilization (1350Kbps/1400Kbps=96.4%) and avoids filling these redundant links, which also verifies the design of the design of Linked Increase algorithm [6].

Table I PTP Link Utilization

| Link | #2⇒#1 | #3⇒#2 | #4⇒#2 | #8⇒#2 | #7⇒#3 | #9⇒#3 | #7⇒#4 | #5⇒#4 |
|---|---|---|---|---|---|---|---|---|
| Bandwidth | 1350 Kbps | 384 Kbps | 581 Kbps | 385 Kbps | 191 Kbps | 193 Kbps | 191 Kbps | 197 Kbps |
| Utilization | 96.4% | 96.0% | 95.1% | 96.3% | 95.9% | 96.5% | 95.5% | 98.5% |
| Link | #11⇒#4 | #5⇒#8 | #6⇒#8 | #12⇒#6 | #11⇒#5 | #10⇒#5 | #10⇒#7 | #9⇒#7 |
| Bandwidth | 193 Kbps | 193 Kbps | 192 Kbps | 198 Kbps | 197 Kbps | 195 Kbps | 193 Kbps | 192 Kbps |
| Utilization | 96.5% | 96.5% | 96.0% | 99.0% | 98.5% | 97.5% | 96.5% | 96.0% |

# 7. Related Work

The NDN transport protocols consist of two functional modules, namely congestion control and adaptive forwarding.

*A. Congestion Control*

The congestion control schemes in NDN broadly fall into two categories: 1) receiver-driven congestion control and 2) hop-by-hop interest shaping. The receiver-driven congestion control scheme in NDN is very similar to that in single-path TCP, as each consumer treats all in-use paths as an entirety (i.e. the forwarding is uncontrollable by consumers). To this end, each consumer adjusts the outgoing traffic/requests to the network according to the congestion signal. The congestion signals mainly include 1) loss, 2) round-trip time (RTT) and 3) explicit congestion notification (ECN). Early studies [23] use timeout to detect the loss. However, as the latencies of paths may be different, the timeout estimation is not reliable [7]. Recent studies mainly focus on the other two types of congestion signals. The RTT-based [2] solution (RAAQM) requires Data packets to record the routing information, such that consumers can distinguish in-use paths and detect congestion for each one according to the variations of RTT. Although RAAQM distinguishes the paths for congestion detection, it cannot control the traffic to each path as in PTP. The ECN-based solutions [7], [18] require the bottleneck routers to explicitly notify consumers of congestion using ECNs. These two types of solutions have their own pros and cons. The RTT-based ones introduce an additional overhead to store the path information, but the ECN-based ones need native supports from routers, which may be difficult in overlay NDN solutions, e.g. congestion does not happen at NDN layer.

---

[5] According to the Max-Flow results

The hop-by-hop interest shaping is a unique research topic in NDN, which prevents the congestion of Data packets by limiting the forwarding rate of Interest packets. Several types of solutions are introduced. Hop-by-hop interest shaping [24], [25] (HbHIS) limits the forwarding rate of Interest packets by explicitly allocating fair bandwidths to different flows. ChoPCoP [7] regulates the forwarding rate based on the number of PI packets and the packets delayed in the link layer. Dynamic Interest Limiting [8] adjusts the forwarding rate based on NACKs from downstream nodes. Hop-by-hop interest shaping [9] (HIS) pre-calculates the optimal interest forwarding rate based on the local and neighbour bandwidths. The hop-by-hop interest shaping can also be used with PTP to 1) detect congestion at the early stage and 2) realize per-flow fairness control even if some consumers are not cooperative.

PTP proposed in this paper does not conflict with hop-by-hop approaches. Particularly, PTP can cooperate with the hop-by-hop traffic control. This is considered in future work.

*B. Adaptive Forwarding*

In hop-by-hop adaptive forwarding area, extensive research works have been proposed to support multipath communications. INFORM [26] is an adaptive hop-by-hop forwarding strategy using reinforcement learning inspired by Q-routing, which discovers temporary copies of content not presented in the routers' forwarding tables. Probability-based Adaptive Forwarding is novel solution inspired by ant colony optimization, which selects the forwarding interfaces based on an RTT distribution. A novel cost-Efficient Multimedia content Delivery approach (EcoMD) [27] is proposed for vehicular networks that leverage the ICN features and a heuristic method to optimize the path selection problem. On-demand Multi-Path Interest Forwarding [28] allocates traffic to disjoint paths via the weighted round-robin scheme based on the round-trip time of each path. Via emulating a liquid piping system, Stochastic Adaptive Forwarding proposed by Daniel *et al.* [19] provides a robust traffic allocation even with incomplete routing information. Carofiglio *et al.* [2] proposed an optimal forwarding strategy, RFA, via solving the multi-flow minimum-cost problem approximately, which uses the number of Pending Interests as the approximate metric (of unbalanced traffic) to balance the congestion on different forwarding interfaces. The core problems of adaptive forwarding are 1) a router cannot get information from the whole network, which causes routers difficulty in making a correct forwarding decision, and 2) routers needs to maintain various states for each flow and for each interface, which reduces the system scalability.

PTP is proposed to offload the adaptive forwarding logic from routers to consumers to improve system scalability. In other words, PTP requires routers to forward the Interest packet according to the specified forwarding interface provided by the tag, the traffic allocation on different paths is actually moved to consumers and become a part of congestion control.

## 8. Conclusion

As NDN brings unique challenges to forwarding design, this paper presented PTP, a multisource/multipath transport protocol to support concurrent multisource/multipath content downloading. In contrast to controlling the traffic to all transmission paths as an entirety, PTP enables independent congestion control on each in-use path. This prevents pulling down the overall throughput by the congestion of a single path. PTP brings immediate and practical benefits: 1) it maximizes resource utilization even though the in-use paths are heterogeneous; 2) the MPTCP-based control law and RTT-compensation ensures fairness among heterogeneous sub-flows and 3) the proposed congestion detection scheme can support in-network caching and traffic aggregations in NDN. PTP develops a novel congestion detection method to detect congestion happening anywhere along the transmission path. The complete protocol stack is implemented by modifying the NDN Forwarding Daemon and evaluated using ndnSIM. The results show that the proposed path-specified traffic control design improves consumers' downloading rate and guarantees fairness among users.

PTP's congestion control scheme is loss-based, which is not reliable in lossy links and is sluggish to congestions. In future work, we plan to evaluate different MPTCP control laws and investigate RTT-based (e.g. Vegas [29]) and congestion-based (e.g. BBR [30]) congestion detection schemes to improve PTP. The optimal path selection strategy is another open question derived from PTP. In future work, we plan to design and evaluate different path selection strategies to optimize QoE for different types of applications. In addition, as FAB is a generic approach to bypass LPM, it will be integrated into the convention NDN approaches for performance evaluations in the future.

## *Acknowledgements*

This publication has emanated from research supported by research grants from Science Foundation Ireland (SFI) under grant number 13/SIRG/2178, with assistance from Chinese 973 Programme grant no. 2013CB329106, Chinese NSF Nos.91230109 & 91430214, and Enterprise Ireland (EI) under the COMAND Technology Gateway programme.

## *Appendix*

In addition to the bandwidth–based section strategy, some other possible strategies are listed here. These different strategies target to optimize the transmission performances from different perspectives.

# Random Selection

A straightforward strategy is to randomly choose $N$ in-use paths from all eligible transmission paths. However, this solution may cause low downloading bandwidth and unstable/high round-trip times, which is ineffective for the vast majority of applications as the consumer may select the low-performance paths accidentally.

# Hop-based Selection (Shortest Path Strategy)

The hop-based selection prefers to choose the in-use paths with smaller numbers of hops. This strategy reduces the total number of hops to deliver different content object, which saves the consumption of bandwidth resources. There are two issues that correspond to the hop-based selection. First, NDN can be an overlay protocol, thus the number of hops measured at NDN layer may not reflect the number of hops at lower layers. Second, if in-use paths share the same bottleneck, the downloading bandwidth will degenerate significantly.

# Latency-based Selection

The latency-based selection prefers the in-use paths with lower latencies. Particularly, the consumer measures the averaged latency (from the RTTs of the probing packets) of each path and sorts them from low to high, then the *N* paths with the lowest latencies are selected. Additionally, the consumer can always replace an in-use path with a higher latency with an unused path with a lower latency after the initial selection. This strategy is expected to be valuable for some real-time communication applications (e.g. VOIP) that are sensitive to high RTTs.

# Latency-Variance-based Selection

Similar to the Latency-based Selection, another option is to select the paths with averaged lower latency variance. Specifically, the consumer sorts the averaged RTTs (from the probing packets) for all paths (include the in-use and alternative paths). Then, a nonlinear function (with a window length *W*) is then applied to the sorted RTTs to obtain the moving variance $\sigma(n)$ of the RTTs.

$$\sigma(n) = \sum_{k=n}^{n+W-1}\left(rtt_k - \frac{\sum_{l=n}^{n+W-1}(rtt_l)}{10}\right)^2 \tag{5}$$

Then the in-use paths are selected from the *N* paths with the lowest variance. This strategy may be valuable for some real-time streaming applications (e.g. Video on Demand) that are sensitive to the jitter of RTTs.

# Hybrid combinations

It is possible to combine the different metrics to compose the more complex strategies. For example, for some specific applications such as the adaptive streaming, which requires both the bandwidth and the latency are important, the strategy can first find out the paths that can satisfy the bandwidth requirement and then select the *N* paths that can minimize the latency jitter. However, as optimizing QoE from different perspectives is out of this paper's scope, the details will not be further discussed.